\documentclass[a4paper, twoside, reqno, 12pt, dvips]{amsart}
\RequirePackage{fix-cm} 

\usepackage{fixltx2e}     

\usepackage{eucal}
\usepackage{xspace}
\usepackage{amsgen}
\usepackage{amssymb}
\usepackage{amsmath}
\usepackage{amsfonts}
\usepackage{MnSymbol}
\usepackage[nice]{nicefrac}
\usepackage{mathrsfs, dsfont}

\usepackage{a4wide}

\usepackage{acronym}

\usepackage{indentfirst}
\usepackage{graphicx}
\usepackage{psfrag}

\usepackage[usenames, dvipsnames, pdf]{pstricks}
\usepackage{epsfig}
\usepackage{pst-grad} 
\usepackage{pst-plot} 

\usepackage{subfigure}

\usepackage{longtable}

\usepackage[french,english]{babel}
\usepackage[latin1]{inputenc}

\usepackage{color}
\definecolor{oneblue}{rgb}{0,0.0,0.75}
\usepackage[colorlinks,
            urlcolor=oneblue,
            linkcolor=oneblue,
            citecolor=oneblue,
            bookmarksopen=false,
            pagebackref]{hyperref}

\numberwithin{equation}{section}

\newcommand{\Ls}{\mathsf{L}}
\newcommand{\N}{\mathcal{N}}

\newcommand{\eps}{\varepsilon}
\renewcommand{\L}{\mathcal{L}}
\renewcommand{\O}{\mathcal{O}}

\newcommand{\RE}{\mathrm{Re}}

\newcommand{\p}{\partial}

\newcommand{\half}{{\textstyle{1\over2}}}

\begin{document}

\title[Vortexons in axisymmetric pipe flows]%
{Vortexons in axisymmetric Poiseuille pipe flows}

\author[F.~Fedele]{Francesco Fedele$^*$}
\address{School of Civil and Environmental Engineering \& School of Electrical and Computer Engineering, Georgia Institute of Technology, Atlanta, USA}
\email{fedele@gatech.edu}
\urladdr{http://savannah.gatech.edu/people/ffedele/Research/}
\thanks{$^*$ Corresponding author}

\author[D.~Dutykh]{Denys Dutykh}
\address{University College Dublin, School of Mathematical Sciences, Belfield, Dublin 4, Ireland \and LAMA, UMR 5127 CNRS, Universit\'e de Savoie, Campus Scientifique, 73376 Le Bourget-du-Lac Cedex, France}
\email{Denys.Dutykh@ucd.ie}
\urladdr{http://www.denys-dutykh.com/}

\begin{abstract}
We present a study on the nonlinear dynamics of small long-wave disturbances to the laminar state in non-rotating axisymmetric Poiseuille pipe flows. At high Reynolds numbers, the associated Navier-Stokes equations can be reduced to a set of coupled Korteweg--de Vries-type (KdV) equations that support inviscid and smooth travelling waves numerically computed using the Petviashvili method. In physical space they correspond to localized toroidal vortices  concentrated near the pipe boundaries (\emph{wall vortexons}) or that wrap around the pipe axis (\emph{centre vortexons}), in agreement with the analytical soliton solutions derived by Fedele (2012) \cite{Fedele2012b}. The KdV dynamics of a perturbation is also investigated by means of an high accurate Fourier-based numerical scheme. We observe that an initial vortical patch splits into a centre vortexon radiating patches of vorticity near the wall. These can undergo further splitting leading to a proliferation of centre vortexons that eventually decay due to viscous effects. The splitting process originates from a radial flux of azimuthal vorticity from the wall to the pipe axis in agreement with the inverse cascade of cross-stream vorticity identified in channel flows by Eyink (2008) \cite{Eyink2008}. The inviscid vortexon most likely is unstable to non-axisymmetric disturbances and may be a precursor to puffs and slug flow formation.
\end{abstract}


\maketitle

\tableofcontents

\section{Introduction}

The laminar Hagen--Poiseuille flow is believed to be linearly stable to periodic or localized infinitesimal disturbances for all Reynolds numbers $\RE$ (see, for example, \cite{Drazin2004}), in contrast to rotating pipe flows that exhibit supercritical bifurcation \cite{Mackrodt1976}. Transition to turbulence in non-rotating pipe flows is thus triggered by finite-amplitude perturbations \cite{Hof2003}. The coherent structures observed at the transitional stage are in the form of vortical patches known as puffs and slugs \cite{Wygnanski1973, Wygnanski1975} that are triggered by non-axisymmetric perturbations, although strong axisymmetric input can lead to similar turbulent structures \cite{Leite1959, Fox1968}. Puffs are spots of vorticity localized near the pipe axis surrounded by laminar flow. Slugs develop along the streamwise direction, while expanding through the entire cross-section of the pipe, and they are concentrated near the wall. Recent theoretical studies tried to relate slug flows to quasi inviscid solutions of the Navier--Stokes (NS) equations for non-rotating pipe flows. In particular, for non-axisymmetric flows Smith and Bodonyi (1982) \cite{Smith1982} revealed the existence of nonlinear neutral structures localized near the pipe axis (centre modes). They are inviscid travelling waves of small but finite amplitude, and unstable equilibrium states (see \cite{Walton2005}). More recently, Walton (2011) \cite{Walton2011} investigated the nonlinear stability of impulsively started pipe flows to axisymmetric perturbations at high Reynolds numbers and he found the axisymmetric analogue of inviscid centre modes. Walton's neutral mode and the inviscid axisymmetric 'slug' structure proposed by Smith \emph{et al}. (1990) \cite{Smith1990} are similar to the slugs of vorticity that have been observed in both experiments \cite{Wygnanski1973} and numerical simulations \cite{Willis2009}. These inviscid structures may play a role in pipe flow transition as precursors to puffs and slugs, since most likely they are unstable to non-axisymmetric disturbances, as the centre modes (see \cite{Walton2005}).

Recently Fedele (2012) \cite{Fedele2012b} investigated the dynamics of non-rotating axisymmetric pipe flows in terms of solitons and travelling waves of nonlinear wave equations. He showed that, at high Reynolds numbers, the dynamics of small and long-wave perturbations to the laminar flow obeys a coupled system of nonlinear Korteweg--de Vries-type (KdV) equations. These set of equations generalize the one-component KdV model derived by Leibovich \cite{Leibovich1968} to study propagation of waves along the core of concentrated vortex flows (see also Benney 1966) and vortex breakdown \cite{Leibovich1984}. Fedele's coupled KdV equations support inviscid soliton and periodic wave solutions in the form of toroidal vortex tubes, hereafter referred to as \emph{vortexons}, which are similar to the inviscid nonlinear neutral centre modes found by Walton (2011) \cite{Walton2011}.

In light of these studies, a generic Navier--Stokes flow can thus be interpreted as a nonlinear sea state \cite{Fedele2008a, Fedele2009} of interacting coherent wave structures of soliton bearing equations. Such sea states may, for example, explain the occurrence of steady 'puffs' observed in both numerical simulations and experiments of turbulent pipe flows \cite{Willis2008a, Willis2009}. Indeed, puffs appear to be dynamically similar to solitons. As turbulence becomes self-sustained at $\RE \approx 2000$, puffs appear at $\RE \approx 2250$ and as $\RE$ increases they slowly delocalize by splitting into two or more puffs, eventually expanding in to slug flow (\cite{Avila2011}). Similarly, a soliton loses energy as it interacts with the background or other solitons, and it delocalizes in space by splitting into many other smaller solitons leading to a solitonic sea state. Furthermore, the 'edge states' computed in numerical experiments of pipe flows at $\RE \approx 10^5$ \cite{Duguet2008} can be interpreted as 'semi-stable' localized travelling waves that neither relaminarize nor become turbulent. Similar studies for Blasius flows have shown that, at high Reynolds numbers, the dynamics is described by a Benjamin--Davis--Acrivos (BDA) integro-differential equation \cite{Ryzhov2010}. This supports soliton structures that explain the formation of spikes observed in boundary-layer transition \cite{Kachanov1993}.

In this paper, we confirm the validity of the theoretical vortexons derived by Fedele (2012) \cite{Fedele2012b}. To do so, we numerically compute traveling waves of the abovementioned KdV equations using the Petviashvili method \cite{Petviashvili1976, Yang2010}) and discuss the associated vortical flow structures. Finally, the numerical investigation of the KdV evolution of a perturbation reveals new dynamical flow features that are physically interpreted and discussed.

\section{Coupled KdV-type equations}

Consider the axisymmetric flow of an incompressible fluid in a pipe of circular cross section of radius $R$ driven by an imposed uniform pressure gradient. The time, radial and streamwise lengths as well as velocities are rescaled with $T$, $R$ and $U_0$ respectively. Here, $T = R/U_0$ is a convective time scale and $U_0$ is the maximum laminar flow velocity. Define a cylindrical coordinate system $(z, r, \theta)$ with the $z$-axis along the streamwise direction, and $(u = -\frac{1}{r}\p_z\Psi, w = \frac{1}{r}\p_r\Psi )$ as the radial and streamwise velocity components of a divergence-free axisymmetric velocity field given in terms of a Stokes streamfunction $\Psi (r,z,t)$. Consider a perturbation to the laminar base flow $W_0(r) = 1 - r^2$ and decompose $\Psi$ as
\begin{equation*}
  \Psi = \Psi_0 + \psi,
\end{equation*}
where $\Psi_0 = \half r^2\left(1 - \half r^2\right)$ represents the stream function of the laminar flow $W_0$, and $\psi$ that of the disturbance. The curl of the NS equations yields the following nonlinear equation for $\psi$ \cite{Itoh1977}
\begin{equation}\label{eq:fg}
  \p_t\Ls\psi + W_0\p_z\Ls\psi - \frac{1}{\RE}\Ls^2\psi = \N(\psi),
\end{equation}
where the nonlinear differential operator
\begin{equation*}
\N(\psi) = -\frac{1}{r}\p_r\psi\p_z\Ls\psi + \frac{1}{r}\p_z\psi\p_r\Ls\psi - \frac{2}{r^2}\p_z\psi\Ls\psi,
\end{equation*}
the linear operator
\begin{equation*}
\Ls = \L + \p_{zz}, \qquad \L = \p_{rr} - \frac{1}{r}\p_r = r\p_r\left(\frac{1}{r}\p_r\right),
\end{equation*}
and $\RE$ is the Reynolds number based on $U_0$ and $R$. The boundary conditions for \eqref{eq:fg} reflect the boundedness of the flow at the centerline of the pipe and the no-slip condition at the wall, that is $\p_r\psi = \p_z\psi = 0$ at $r = 1$.

Drawing from Fedele (2012) \cite{Fedele2012b}, the solution of \eqref{eq:fg} can be given in terms of a complete set of orthonormal basis $\{\phi_j(r)\}$ as
\begin{equation}\label{eq:ex1}
  \psi(r,z,t) = \sum_{j=1}^\infty \phi_j(r) B_j(z,t),
\end{equation}
where $B_j$ is the amplitude of the radial eigenfunctions $\phi_j$ that satisfy the Boundary Value Problem (BVP) (see \cite{Fedele2005})
\begin{equation*}
  \L^2\phi_j = -\lambda_j^2\L\phi_j,
\end{equation*}
with $\frac{1}{r}\phi_j$ and $\frac{1}{r}\p_r\phi_j$ bounded at $r \to +0$, and $\phi_j = \p_r\phi_j = 0$ at $r = 1$. Since $\phi_j$ satisfies the pipe flow boundary conditions \emph{a priori}, so does $\psi$ of \eqref{eq:ex1}. Note that even for a finite number $N$ of modes, the vorticity of the velocity field associated to the truncated expansion for $\psi$ is divergence-free. The positive eigenvalues $\lambda_j$ are the roots of $J_2(\lambda_j) = 0$, where $J_2(r)$ is the Bessel function of first kind of second order (see \cite{Abramowitz1965}). For example, for the first two least stable eigenmodes $\lambda_1 \approx 5.136$ and $\lambda_2 \approx 8.417$, respectively. A Galerkin projection of \eqref{eq:fg} onto the Hilbert space spanned by $\{\phi_j\}_{j=1}^\infty$ yields an infinite set of coupled generalized Camassa--Holm (CH) equations \cite{Camassa1993}
\begin{equation}\label{eq:CH}
  \p_t B_j + c_{jm}\p_z B_m + \beta_{jm}\p_{zzz}B_m + \alpha_{jm}\p_{zzt} B_m + N_{jnm}(B_n, B_m) + \frac{\lambda_j^2 B_j}{\RE} = 0,
\end{equation}
where
\begin{equation}\label{eq:N}
  N_{jnm}(B_n, B_m) = F_{jnm}B_n\p_z B_m + \\ G_{jnm}\p_z B_n\p_{zz}B_m + H_{jnm}B_n\p_{zzz}B_m,
\end{equation}
and summation over repeated indices $n$ and $m$ is implicitly assumed. The tensors $c_{jm}$, $\beta_{jm}$ and $\alpha_{jm}$ of the advection and dispersive terms are function of the base flow $W_0$. These and $F_{jnm}$, $G_{jnm}$, $H_{jnm}$ are given in Fedele (2012) \cite{Fedele2012b}.  In simple words, the equations \eqref{eq:CH} govern the nonlinear interaction of the radial structures $\{\phi_j(r)\}$ that are advected and dispersed in the streamwise direction by the laminar flow. Note that CH type equations arise also as a regularized model of the 3-D NS equations, the so called Navier--Stokes-alpha model \cite{Foias2001}.

Starting from \eqref{eq:CH} Fedele (2012) \cite{Fedele2012b} showed that the nonlinear dynamics of a small long-wave perturbation can be reduced to that on the 'slow manifold' of the laminar state spanned by the first few $N$ least stable modes $\phi_j$. The higher damped modes ($j > N$) can be legitimately set as slaved and so neglected as long as (\textit{i}) the amplitudes $B_j$ remain for all time in a small neighborhood of zero (the laminar state), and (\textit{ii}) the non-resonant condition $\lambda_{i_1}^2 + \lambda_{i_2}^2 + \ldots + \lambda_{i_n}^2 \neq \lambda_j^2$ holds true, with $\{i_1, i_2, \ldots, i_n\}$ any permutation of size $n \leq N$ drawn from the set $j = 1,\ldots, N$ (see \cite{DelaLlave1997}). For the BVP of \eqref{eq:ex1} (ii) has been verified to hold numerically up to $N \sim 10^4$. As a result, in the long-wave limit higher order nonlinearities in \eqref{eq:N} can be neglected and \eqref{eq:CH} reduces to the coupled damped KdV equations
\begin{equation}\label{eq:main2}
\p_t B_j + c_{jm}\p_z B_m + \beta_{jm}\p_{zzz}B_m + \\ F_{jnm} B_n\p_z B_m +\frac{\lambda_j^2 B_j}{\RE} = 0,
\end{equation}
where $j = 1,\ldots,N$. In particular, Fedele (2012) \cite{Fedele2012b} proved that if $B_j = \eps b_j(\xi, \tau)$, with $\eps \sim \O(\RE^{-\frac{2}{5}})$ and $\xi = \eps^{\frac{1}{2}}(z - Vt)$, $\tau = \eps^{\frac{3}{2}} t$, then the dynamics is primilary inviscid for time scales much less than $t \sim \O(\eps^{-\frac{5}{2}}) = \O(\RE^{6.25})$, and $V$ is given approximately by the average of the eigenvalues of $c_{jm}$. Moreover, the inviscid KdV system \eqref{eq:main2} supports analytical travelling waves (TWs). In physical space, they correspond to localized or periodic toroidal vortices, which travel slightly slower than the maximum laminar flow speed $U_0$, i.e. $V \simeq 0.77$. For $N = 2$, Fedele (2012) \cite{Fedele2012b} showed that the vortical structures are localized near the wall (wall vortexon, $B_1$ and $B_2$ have the same sign) or wrap around the pipe axis (centre vortexon, $B_1$ and $B_2$ have opposite signs). They have a non-zero streamwise mean, but they radially average to zero to conserve mass flux through the pipe. The wall vortexon is the axisymmetric analogue of the 2-D rolls+streaks stage of the self-sustaining process (SSP) of Waleffe \cite{Waleffe1995, Waleffe1995a, Waleffe1997} in Couette flows (see also \cite{Gibson2008}) or that of Wedin and Kerswell (2004) \cite{Wedin2004} in pipe flows (see also Hof et al. 2004). Instead, centre vortexons may be related to the inviscid neutral axisymmetric slug structures discovered by Walton (2011) in unsteady pipe flows, which are similar to the centre modes proposed by Smith \emph{et al}. (1990) \cite{Smith1990}.

In the following we will confirm Fedele's theoretical results by computing numerically TWs of the inviscid form of the KdV equations \eqref{eq:main2}.

\section{Cnoidal waves and vortexons}

Consider the ansatz $B_j = q_j + F_j(z-ct)$, where $q_j$ are free parameters and $c$ is the velocity of the TW (here both $c$ an $q_j$ are normalized with respect to the maximum laminar velocity $U_0$). We make use of the Petviashvili method \cite{Petviashvili1976, Yang2010} to solve numerically the nonlinear steady problem for $F_j$ (in the moving frame $z-ct$). This numerical approach has been successfully applied to derive TWs of the spatial Dysthe equation \cite{Fedele2011} and the compact Zakharov equation for water waves \cite{Fedele2012a}.

\begin{figure}
  \centering
  \includegraphics[trim=0.0cm 0.0cm 0.0cm 0.0cm, clip=true, width=0.95\textwidth]{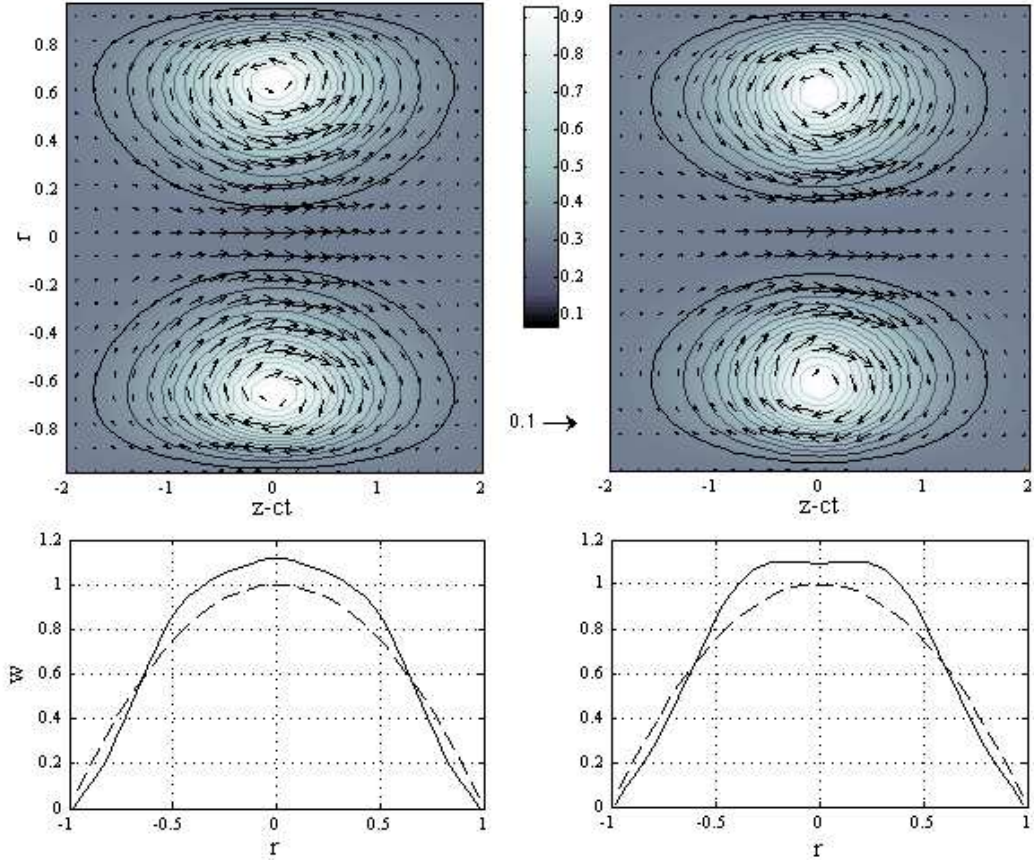}
  \caption{Wall vortexon: (top-left) streamlines of the three-component KdV solution and (bottom-left) velocity profiles of the perturbed (solid) and laminar (dash) flows; (right) two-component KdV counterpart ($c=0.68$, $q=0.1$).}
  \label{fig:fig1}
\end{figure}

\begin{figure}
  \centering
  \includegraphics[trim=0cm 0cm 0cm 0cm, clip=true, width=0.85\textwidth]{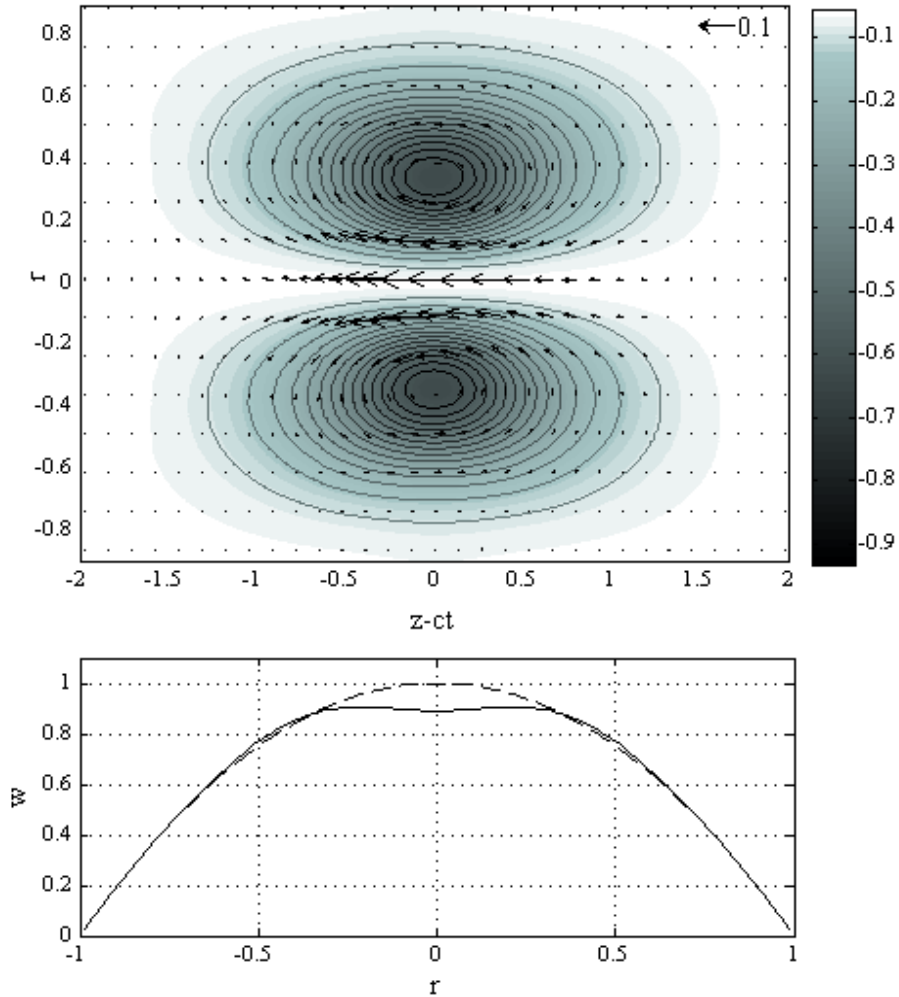}
  \caption{Centre vortexon: (top) streamlines of the three-component KdV
solution and (bottom) velocity profiles of the perturbed (solid) and
laminar (dash) flows ($c=0.62$, $q=0$).}
  \label{fig:fig2}
\end{figure}

The Petviashvili method converged to both wall and centre localized TWs (solitons or solitary waves) in the range of parameters $c \sim [0.62, 0.68]$ and $q_j = q \sim [0, 0.9]$. The generic topology of the flow structure of the TW disturbance is the same as that of the theoretical counterpart derived by Fedele (2012) \cite{Fedele2012b}: toroidal tubes of vorticity localized near the pipe boundaries (wall vortexon, see Figure~\ref{fig:fig1} ) or that wrap around the pipe axis (centre vortexon, Figure~\ref{fig:fig2}). In particular, the two-component wall vortexon is compared against the three-component counterpart in Figure~\ref{fig:fig1}, which reports the streamlines of the perturbations. As one can see, the topology of the two vortical structures is similar.

It appears that as $N$ increases the effects of the higher modes tend to somewhat diminish, but a rigorous proof of this statement is far beyond the scope of this work. We just point out that the proper theoretical framework for such a proof is provided by slow-manifold type theorems (see, for example, \cite{DelaLlave1997}).

\section{Vortexon dynamics}

\begin{figure}
  \centering
  \includegraphics[trim=1.5cm 1.5cm 1.5cm 1cm, clip=true, width=0.93\textwidth]{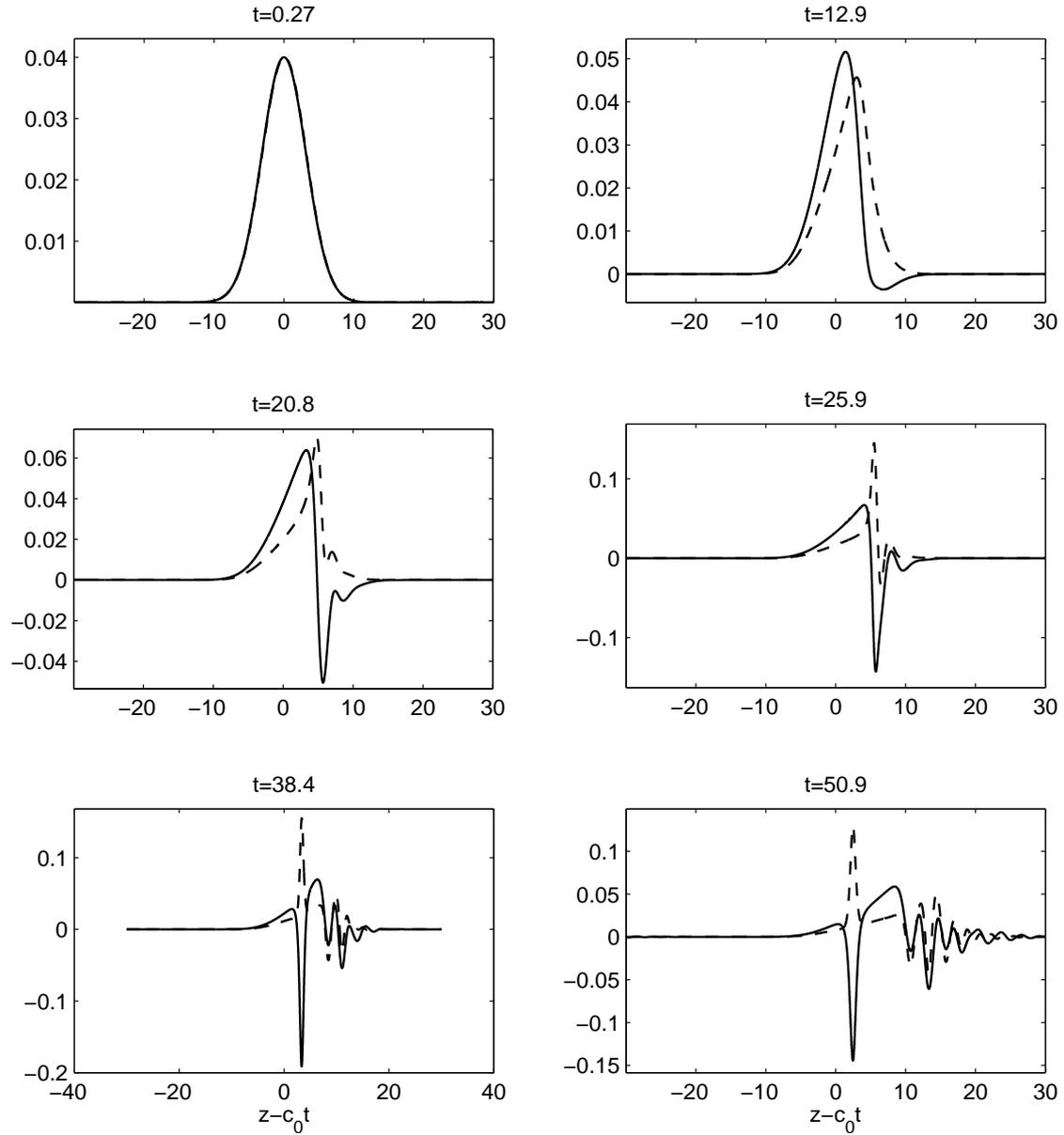}
  \caption{KdV dynamics of a perturbation: wave components $B_1$ (solid) and $B_2$ (dash) at different instants of times ($\RE=5000$, speed of the reference frame $c_0=0.62$).}
  \label{fig:fig3}
\end{figure}

Hereafter, we investigate the dynamical evolution of a localized disturbance under the two-component KdV dynamics with dissipation (see, equation \eqref{eq:main2}). To do so, we exploit a highly accurate Fourier-type pseudo-spectral method that is described in Fedele \& Dutykh (2012) \cite{Fedele2012a}. Figure \ref{fig:fig3} depicts snapshots of the two-component KdV solution at different times and the streamlines of the associated vortical structures are shown in\ Fig. 4, ($\RE \sim 5000$). The waveform of each component initially steepens up to a limiting shape and then breaks into a soliton and radiative waves as a result of the competition between the laminar-flow-induced wave dispersion and the nonlinear energy cascade associated to the KdV nonlinearities. In physical space we observe a compression of the initial vortical structure that splits in to a centre vortexon and patches of vorticity in the form of weak wall vortexons. On the simulated time scales, the flow evolution is basically inviscid, but it eventually decays on the long time scale $t \sim \O(\RE^{6.25})$, due to viscous effects (see \cite{Fedele2012b}). The centre vortexon arises due to a radial flux $F_{\theta r}^{(\omega)} \sim u \omega_{\theta}$ of azimuthal vorticity $\omega_{\theta} = -\frac{1}{r}\Ls\psi$ from the wall to the pipe axis. This is the 'inverse cascade' mechanism of cross-stream vorticity identified by Eyink (2008) \cite{Eyink2008} in channel flows. The patches of radiated near-wall vorticity may further split causing a proliferation of centre and wall vortexons until viscous effects attenuate and annihilate the splitting process. The formation of a '\emph{vortexon slug}' is clearly seen in Figure \ref{fig:fig5}, which reports the space-time plot of the difference $\beta = |B_1 - B_2|$ of the two wave components. Here, larger values of $\beta$ trace centre vortexons ($B_1$ and $B_2$ have opposite sign), whereas smaller values of $\beta$ are associated to wall vortexons ($B_1$ and $B_2$ have the same sign). The streamlines of the associated flow perturbation at different times are shown in Figure~\ref{fig:fig6}. Note that the evolution is similar to that of puff splitting observed in the spreading of pipe turbulence at transition \cite{Avila2011}. Here, turbulence arises as a competition between puff decay (death) and puff splitting (birth) processes. A turbulence state can persist only when new puffs are produced faster than their decay. Instead, the proliferation of centre vortexons is mostly inviscid and due to the competition between dispersion and steepening of radial structures (the eigenmodes $\phi_j(r)$) that are advected and dispersed in the streawise direction by the laminar flow.  Clearly, it is a bit of a stretch to identify vortexon slugs as the realistic ones observed in experiments, which also have a substantial non-axisymmetric component. However, similarly to the inviscid neutral modes found by Walton (2011) \cite{Walton2011}, centre vortexons most likely are unstable to non-axisymmetric disturbances, and may persist viscous attenuation as precursors to puffs and slugs.

\begin{figure}
  \centering
  \includegraphics[trim=0cm 0cm 0cm 0cm, clip=true, width=0.95\textwidth]{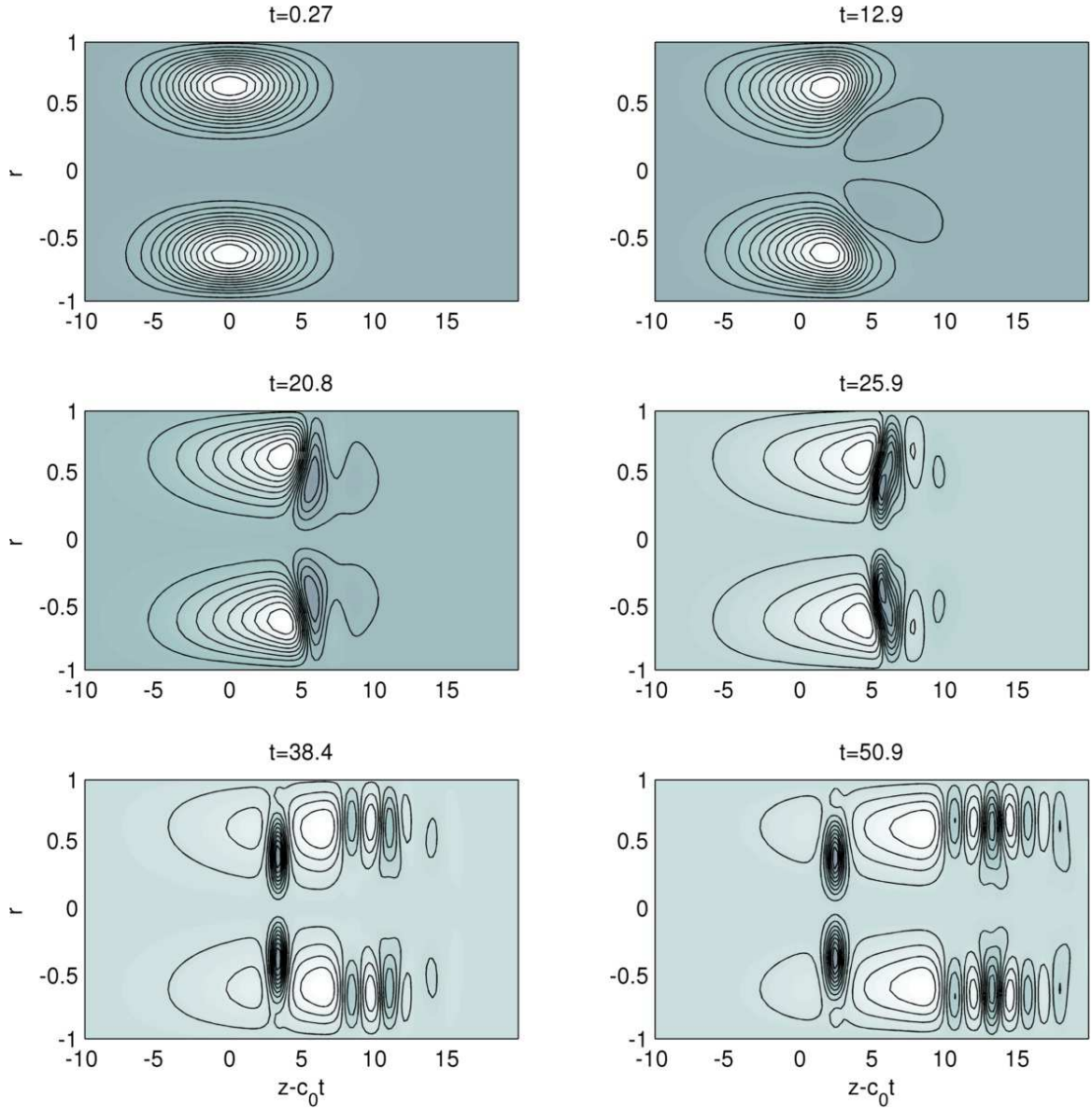}
  \caption{KdV dynamics of a perturbation: streamlines of the vortical structures associated to the wave components $B_1$ and $B_2$ in Fig.~\ref{fig:fig3}.}
  \label{fig:fig4}
\end{figure}

\begin{figure}
  \centering
  \includegraphics[trim=1.3cm 1.1cm 0.9cm 1.2cm, clip=true, width=0.89\textwidth]{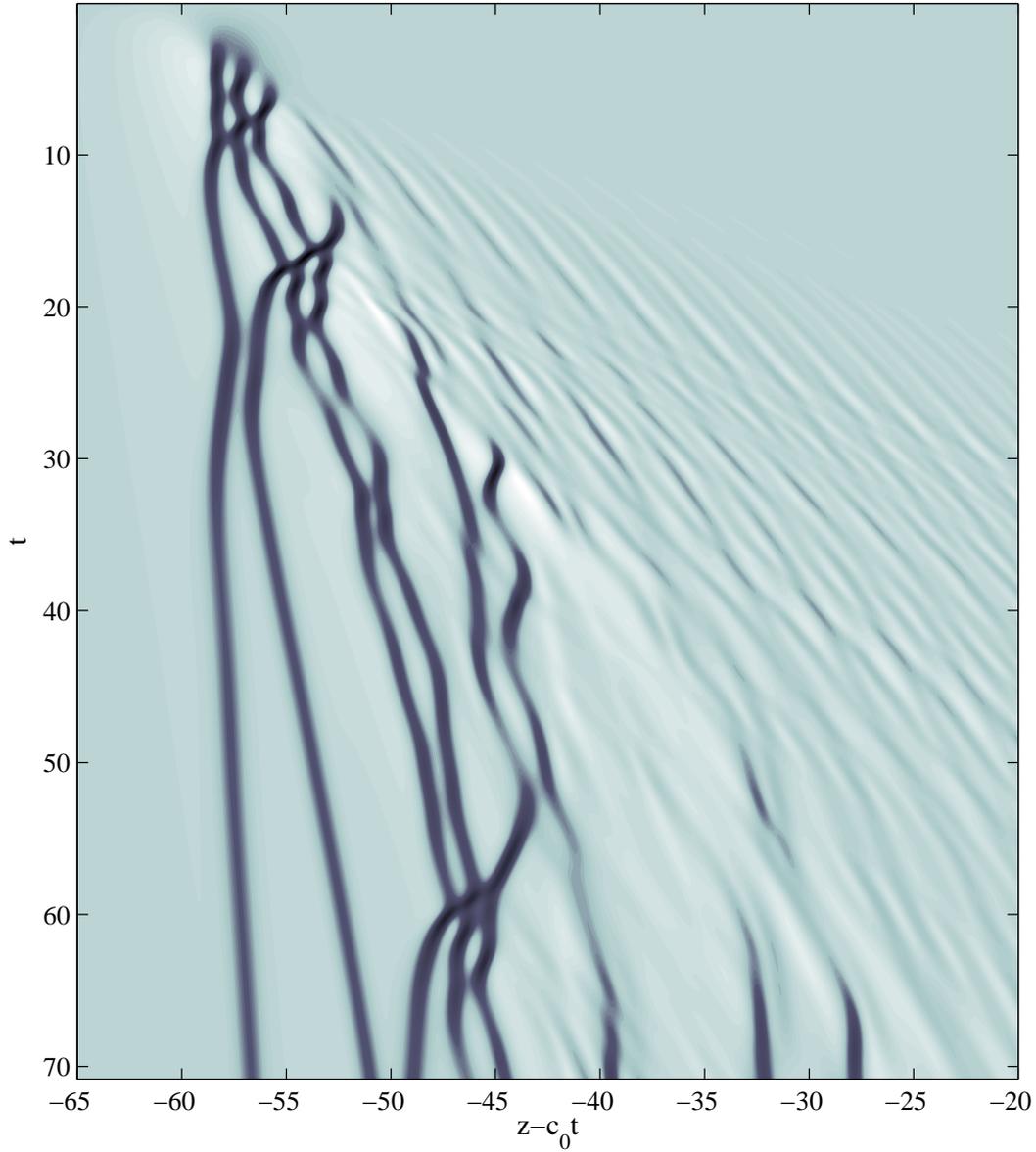}
  \caption{Space-time evolution of $\beta = |B_1 - B_2|$ for $\RE=5000$, speed of the reference frame $c_0=0.62$. Large values of $\beta$ trace centre vortexons ($B_1$ and $B_2$ have opposite sign), whereas smaller values are associated to wall vortexons ($B_1$ and $B_2$ have same sign).}
  \label{fig:fig5}
\end{figure}

\begin{figure}
  \centering
  \includegraphics[trim=1.6cm 0.7cm 1.5cm 1.0cm, clip=true, width=0.96\textwidth]{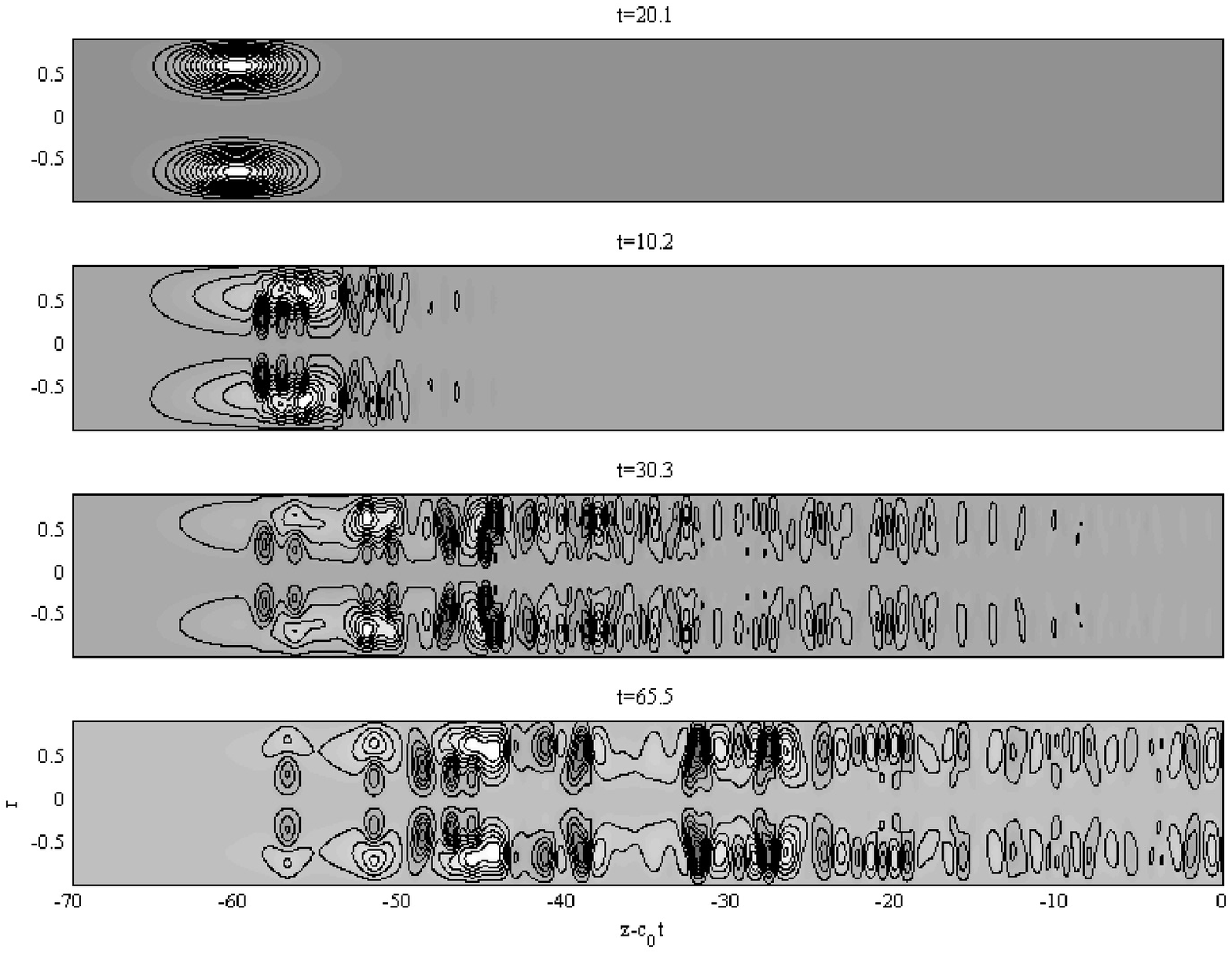}
  \caption{Long-time evolution of a perturbation: streamlines of the vortical structures associated to the wave dynamics of Fig.~\ref{fig:fig5}. The snapshot times from top to bottom are $t = 10.2$, $20.1$, $30.3$ and $65.5$ respectively.}
  \label{fig:fig6}
\end{figure}

\section{Conclusions}

In the limit of small and long-wave disturbances the axisymmetric Navier--Stokes equations for non-rotating pipe flows reduce to coupled KdV equations. In the inviscid limit they support analytical cnoidal solitary waves as proven by Fedele \cite{Fedele2012}. These theoretical results are in agreement with numerical solutions of TWs constructed using the Petviashvili method. The associated flow structures are toroidal vortices that travel slightly slower than the maximum laminar flow speed, and they can be localized near the wall (wall vortexon) or wrap around the pipe axis (centre vortexon).

We have also investigated numerically the evolution of a perturbation in accord to the KdV dynamics. We found that it evolves into a 'slug' of centre vortexons that split from patches of near-wall vorticity due to a radial flux of azimuthal vorticity from the wall to the pipe axis in agreement with Eyink (2008) \cite{Eyink2008}. The centre vortexon is similar to the inviscid neutral mode found by Walton (2011) \cite{Walton2011}, and it could be a precursor to puffs and may relate to the formation of a slug structure in pipe flows.

Finally, we wish to emphasize the relevance of this work to the understanding of transition to turbulence. For chaotic dynamical systems the periodic orbit theory (POT) in Cvitanovi\'c and Eckhardt (1991) \cite{Cvitanovic1991} and Cvitanovi\'c (1995) \cite{Cvitanovic1995} interpret the turbulent motion as an effective random walk in state space where chaotic (turbulent) trajectories visit the neighborhoods of equilibria, travelling waves, or periodic orbits of the NS equations, jumping from one saddle to the other through their stable and unstable manifolds \cite{Wedin2004, Kerswell2005, KerswellL2007, Gibson2008}. Non-rotating axisymmetric pipe flows do not exibhit chaotic behaviour (see, e.g., \cite{Patera2006, Willis2008a}), and so the associated KdV equations (even with dissipation). However, forced and damped KdV equations are chaotic and the attractor is of finite dimension (see, for example, \cite{Cox1986, Grimshaw1994}). Thus, the study of the reduced KdV equations associated to forced axisymmetric Navier--Stokes equations using POT may provide new insights into the nature of slug flows and their formation.

\section*{Acknowledgements}

F.~\textsc{Fedele} acknowledges the travel support received by the Geophysical Fluid Dynamics (GFD) Program to attend part of the summer school on ``\textit{Spatially Localized Structures: Theory and Applications}'' at the Woods Hole Oceanographic Institution in August 2012.

D.~\textsc{Dutykh} acknowledges the support from ERC under the research project ERC-2011-AdG 290562-MULTIWAVE.

\bibliography{biblio}
\bibliographystyle{plain}

\end{document}